\documentclass[sigconf]{acmart}

\AtBeginDocument{%
  \providecommand\BibTeX{{%
    \normalfont B\kern-0.5em{\scshape i\kern-0.25em b}\kern-0.8em\TeX}}}

\setcopyright{acmcopyright}
\copyrightyear{2018}
\acmYear{2018}
\acmDOI{XXXXXXX.XXXXXXX}

\acmPrice{15.00}
\acmISBN{978-1-4503-XXXX-X/18/06}
\newcommand{\ie}{\emph{i.e., }}
\newcommand{\eg}{\emph{e.g., }}

\usepackage{booktabs}

\usepackage{arydshln} 

\usepackage{booktabs}
\usepackage{enumitem}
\usepackage{verbatim}
\usepackage[ruled]{algorithm2e}
\usepackage{multirow}
\usepackage{subfigure} 
\usepackage{ragged2e}
\usepackage{caption}
\usepackage{subcaption}
\usepackage{graphicx}
\usepackage[normalem]{ulem}
\useunder{\uline}{\ul}{}
\usepackage{amsmath}
\usepackage{setspace}
\usepackage{bm}
\usepackage{longtable}
\begin{document}

\title{Exact and Efficient Unlearning for Large Language Model-based Recommendation}

\author{Zhiyu Hu}
\affiliation{%
  \institution{University of Science and Technology of China}
    \city{Hefei}
  \country{China}}
\email{zhiyuhu@mail.ustc.edu.cn}

\author{Yang Zhang}
\affiliation{%
  \institution{University of Science and Technology of China}
    \city{Hefei}
  \country{China}}
\email{zy2015@mail.ustc.edu.cn}

\author{Minghao Xiao}
\affiliation{%
  \institution{University of Science and Technology of China}
    \city{Hefei}
  \country{China}}
\email{xiaominghao@mail.ustc.edu.cn}

\author{Wenjie Wang}
\affiliation{%
  \institution{National University of Singapore}
  \country{Singapore}}
\email{wenjiewang96@gmail.com}

\author{Fuli Feng}
\affiliation{%
  \institution{University of Science and Technology of China}
    \city{Hefei}
  \country{China}}
\email{fulifeng93@gmail.com}

\author{Xiangnan He}
\affiliation{%
  \institution{University of Science and Technology of China}
    \city{Hefei}
  \country{China}}
\email{xiangnanhe@gmail.com}
\renewcommand{\shortauthors}{Trovato and Tobin, et al.}

\begin{abstract}
The evolving paradigm of Large Language Model-based Recommendation (LLMRec) customizes Large Language Models (LLMs) through parameter-efficient fine-tuning (PEFT) using recommendation data. The inclusion of user data in LLMs raises privacy concerns. To protect users, the unlearning process in LLMRec, specifically removing unusable data (e.g., historical behaviors) from established LLMRec models, becomes crucial. However, existing unlearning methods are insufficient for the unique characteristics of LLMRec, mainly due to high computational costs or incomplete data erasure.
In this study, we introduce the Adapter Partition and Aggregation (APA) framework for exact and efficient unlearning while maintaining recommendation performance. APA achieves this by establishing distinct adapters for partitioned training data shards and retraining only the adapters impacted by unusable data for unlearning. To preserve recommendation performance and mitigate considerable inference costs, APA employs parameter-level adapter aggregation with sample-adaptive attention for individual testing samples. Extensive experiments substantiate the effectiveness and efficiency of our proposed framework.

\end{abstract}





\maketitle
\section{introduction}
Large language models have demonstrated exceptional capabilities in content comprehension and generation, sparking interest in applying them in Web applications~\cite{llM4med,llM4visual,llM4multimodel,lin2023multi}. Recommender systems, as a primary channel for personalized content distribution, can also benefit from these capabilities in understanding items and users~\cite{collm}, pushing the emergence of large language model-based recommendation paradigm. The current standard approach for specializing LLMs for recommendation is parameter-efficient fine-tuning~\cite{tallrec,bigrec,collm,yuanfj-adapter} using recommendation data. However, incorporating recommendation data (\eg historical behaviors) increases the risk of personal data leakage due to the vulnerability of LLMs~\cite{private1,private2,private3,private4}. To safeguard the privacy of users, particularly vulnerable populations, LLMRec unlearning becomes crucial, which targets at the timely and effective removal of some personal data~\cite{sisa} (termed \textit{unusable data}~\cite{ifru}) from developed LLMRec models.

To the best of our knowledge, there is currently no dedicated research on LLMRec unlearning. Despite the significant advancements of recommendation unlearning for traditional models~\cite{ifru,liu2022forgetting}, these approaches are unfeasible for LLMRec models due to the high computation cost associated with handling billions of model parameters. Some very recent studies~\cite{InContextUnlearn,HarryUnlearning,yao2023large} investigate efficient unlearning techniques for information encoded in a LLM by extending traditional methods~\cite{yao2023large} or utilizing in-context learning~\cite{InContextUnlearn}. However, applying these methods for LLMRec will encounter the risk of incomplete removal due to their approximate nature. In contrast, LLMRec unlearning requires complete removal of the unusable data to comply with relevant regulations such as the General Data Protection Regulation~\cite{regulation2018general}. Additionally, LLMRec unlearning must maintaining maintain overall recommendation performance to ensure a satisfactory user experience.

Achieving desirable LLMRec unlearning hinges on retraining PEFT adapters using data partitioning. Inspired by traditional unlearning methods~\cite{sisa,receraser,graphunlearn}, retraining has proven to be a reliable method for ensuring exact unlearning. By employing a partitioning strategy that divides the training data into disjoint shards and training sub-models for each shard, unlearning efficiency can be maintained as only the sub-models affected by unusable data are retrained. Since personal data is exclusively stored in the PEFT adapter (\eg LoRA~\cite{lora}), retraining adapters on relevant shards incurs relatively low costs. Additionally, the PEFT adapter can quickly learn from a minimal number of examples, enabling further reduction in shard size and retraining costs. Considering these factors, we propose leveraging an adapter partition-empowered retraining approach for LLMRec unlearning.

The partition strategy in LLMRec presents a distinct challenge in terms of inference latency since aggregating prediction results from different adapters is necessary to integrate knowledge for maintaining high recommendation performance~\cite{sisa,receraser,graphunlearn}. However, such aggregation becomes infeasible for LLMRec models due to the computationally expensive nature of LLM inference~\cite{costhigh}. Generating predictions from $K$ adapters would result in a $K$ times increase in the inference cost of LLMs, leading to substantial rises in energy consumption and service latency. For this challenge, we consider adapter aggregation at parameter level to enable a single-pass inference. Additionally, the partition and aggregation should be carefully designed to ensure a recommendation performance comparable to adapter retraining without partitioning~\cite{modelsoup}. 

To this end, we introduce the \textit{Adapter Partition and Aggregation} framework for exact and efficient LLMRec unlearning while maintaining overall recommendation performance. APA trains individual adapters on partitioned training data shards and leverages adapter weight aggregation during inference. As to partition, APA divides the training data into balanced and heterogeneous shards based on semantic characteristics~\cite{bigrec} to facilitate keeping recommendation performance~\cite{modelsoup}. 
As to aggregation, we adopt a sample-adaptive approach for each testing sample that assigns adapter attention based on the performance of adapters on similar validation samples. This prioritizes higher-performing adapters, thereby enhancing overall performance. Notably, additional training is unnecessary for our adaptive aggregation, unlike traditional unlearning methods~\cite{receraser,graphunlearn}, thus avoiding extra unlearning costs.

The main contributions can be summarized as follows:
\begin{itemize}[leftmargin=*]
    \item New problem:  To our knowledge, this is the first study to formulate and explore exact unlearning within the realm of LLMRec.
    \item New technique: We introduce the APA method, an extension of partition-based unlearning, tailored to scenarios where inference involves high computational costs.
    \item Experiments: We conduct extensive experiments on two real-worlds, verifying the effectiveness of the proposal.
\end{itemize}

\begin{figure}
\label{rec-tuning}

\begin{minipage}{0.195\textwidth}
  \includegraphics[width=\linewidth]{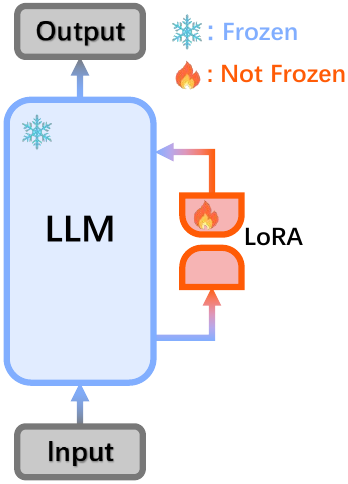}
\end{minipage}
\hfill
\begin{minipage}{0.26\textwidth}
\label{tablea}
\begin{tabular}{@{}ll@{}}
\toprule
\multicolumn{2}{c}{\textbf{Instruction Input}}                                                                                                                                              \\ \midrule
Instr.: & \begin{tabular}[c]{@{}l@{}}Given the user's prefe-\\rence and  unpreference, ...\end{tabular} \\ \midrule
Input:       & \begin{tabular}[c]{@{}l@{}}User preference: ...  \\ User unpreference: ...\\ Target new movie: ...\end{tabular}                            \\ \midrule
\multicolumn{2}{c}{\textbf{Instruction Output}}                                                                                                                                             \\ \midrule
Output:      & Yes.                                                                                                                                                                         \\ \bottomrule
\end{tabular}

\end{minipage}
\caption{The left diagram illustrates the classic structure of LoRA, while the right table provides a sample for recommendation instruction data.}
\label{tableandfig}
\vspace{-15pt}
\end{figure}

\section{Preliminaries}

\begin{figure*}[th]
  \centering
  \includegraphics[width=0.85\linewidth]{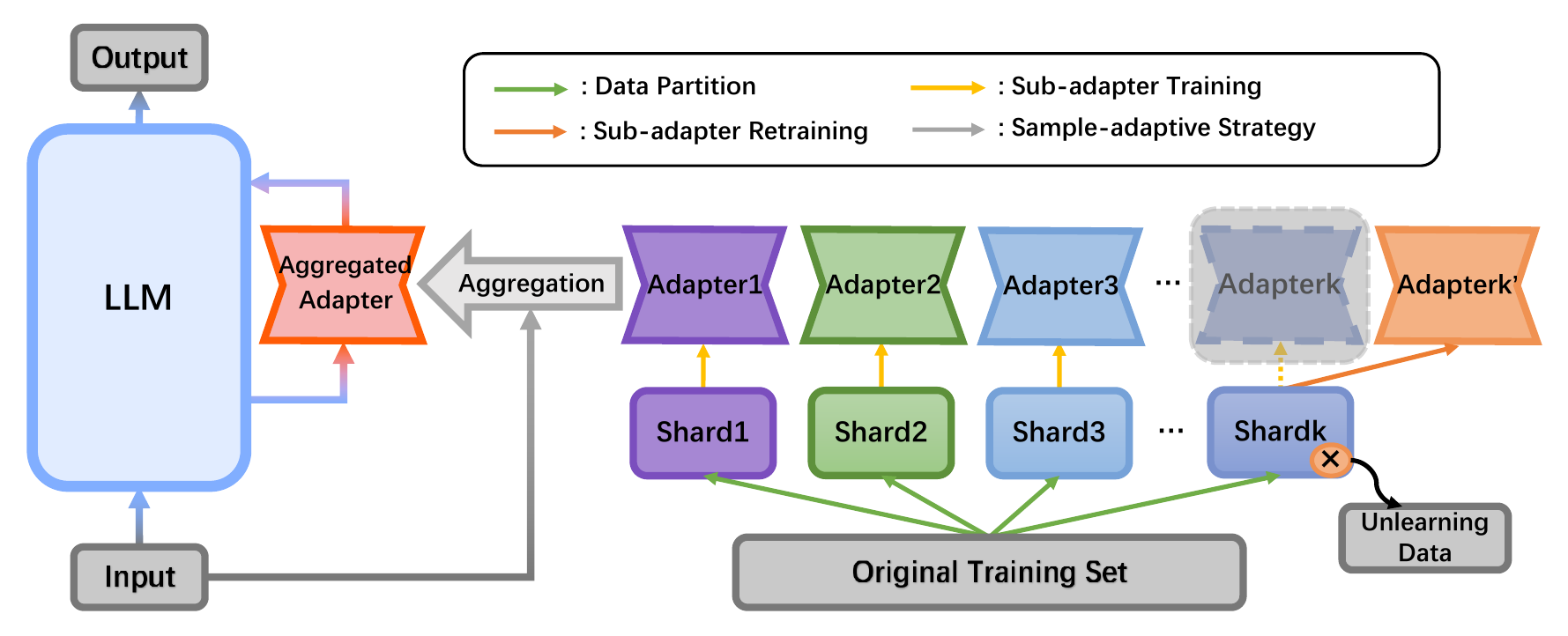}
  \caption{ Illustration of our APA framework, which
consists of three parts: Data Partition, Adpter training,
and Adpter aggregation. When a user requests to erase data $D_r$, only the sub-
LoRA adapters affected by $D_r$ need to be retrained.}
  \label{fig:framework}
   \vspace{-5pt}
\end{figure*}
In this section, we first briefly introduce the prerequisite knowledge of PEFT and LLMRec. Then, we give the problem formulation.

\subsection{PEFT}
Fine-tuning LLMs with domain-specific data is an effective method to tailor LLMs for domain-specific tasks. Given that LLMs typically comprise billions of parameters, full tuning is a resource-intensive and time-consuming process. Recent work~\cite{aghajanyan2020intrinsic} shows that LLMs have a low intrinsic dimension that can match the performance of the full parameter space. PEFT provides a solution to this challenge by keeping the most of model weights frozen and only updating a part of the parameters. These learnable parameters are controlled by an adaptation module (termed adapter). 


\par \textit{LoRA}. 
In this work, we focus on \textit{Low-Rank Adaptation} (LoRA)~\cite{lora}, a prominent and widely adopted PEFT solution.
To make fine-tuning more efficient, LoRA adds pairs of rank-decomposition weight matrices to existing weights of the LLM in a plug-in manner and only trains the newly added weights for learning tasks. The rank-decomposition design would ensure that the addition of weight matrices introduces only a small number of learnable parameters, thereby expediting the fine-tuning process. More specifically, for a matrix multiplication layer within LLMs, a LoRA module adds weight matrices as follows:
\begin{equation}\small\label{eq:lora} 
 o = W_{0}x + BAx,
\end{equation}
where $x$ and $h$ represent the input and the resulting output, respectively. $W_{0} \in \mathcal{R}^{d{1} \times d_{2}}$  denotes the original model weight matrix, while $B\in \mathcal{R}^{d_{1} \times r}$  and $A \in \mathcal{R}^{r \times d_{2}}$ constitute the pair of rank-decomposition weight matrices, with $d_{1}$, $d_{2}$, and $r$ representing the dimensions involved. Notably, $r \ll \min(d_{1},d_{2})$, meaning that the number of parameters introduced by $BA$ is significantly fewer than that of $\mathcal{W}_{0}$ because $ d_{1}r+rd_{2} \ll d_{1}d_{2}$. During the fine-tuning process, only $A$ and $B$ are adjustable. In a similar way, a LoRA module (also called a LoRA adapter) is generally applicable to any LLM layer desired for updating.   

\subsection{LLMRec}
%


PEFT with recommendation data has emerged as a de facto standard to specialize LLMs for recommendation tasks. Numerous studies have appeared and showcased the effectiveness of such LLMRec methods. In our research, we focus on the widely popular LLMRec method called TALLRec~\cite{tallrec}, considering its broad applicability and representation of the general paradigm in the field of LLMRec.




\textit{TALLRec.} TALLRec employs LoRA tuning techniques to align LLMs with the recommendation task using recommendation instruction data. This approach involves the conversion of user-item interaction data into language instructions, as exemplified in Figure~\ref{tableandfig}. Each instruction comprises both an input and an output component. Within the instruction input, TALLRec represents items using their titles and user preferences or non-preferences are conveyed by referencing historical item titles, it also instructs LLMs to respond with either ``Yes'' or ``No'' to indicate the user's preference for a target item. The response is included in the instruction output. With the instruction data, TALLRec performs fine-tuning of the LLM using a LoRA adapter to learn the recommendation task. Let $\mathcal{D}$ represent the set of all converted instruction data for training, and then the optimization problem can be formulated as follows:
\begin{equation}\small \label{eq:tallrec}
\mathop{\max}\limits_{\Phi} \sum_{(x_i,y_i)\in\mathcal{D}}\sum_{t=1}^{|y|}log(P_{\Theta_{0} + \Phi}({y_{i}}_{t}|x_{i},{y_i}_{<t})),
\end{equation}
where $x$ and $y$ represent the instruction input and output of a data sample in $\mathcal{D}$,  $y_{t}$ represents the $t$-th text token of $y$, $y_{<t}$ denotes the text tokens that precede $y_{t}$, and $P_{\Theta_{0} + \Phi}(y_{t}|x,y_{<t}))$ signifies the predictive probability of $y_{t}$ by the LLM. $\Theta_{0}$ refers to the existing parameters of the original LLM, and $\Phi$ encompasses all model parameters within the LoRA adapter, including the $A$ and $B$ as defined in Equation~\eqref{eq:lora} for all layers.  Notably, only the LoRA adapter parameters $\Phi$ would be updated.

\subsection{Problem Formulation}

Let $\mathcal{D}_{-r} \subset \mathcal{D}$ represent the data that a user wishes to remove from a PEFT LLMRec model $f$ that was initially trained with $\mathcal{D}$. Following previous work, we assume the size of $\mathcal{D}_{-r}$ is very small, \eg $|\mathcal{D}_{-r}|=1$.
We try to obtain a retrained model using only the remaining data, denoted as $\mathcal{D}_{r} = \mathcal{D} - \mathcal{D}_{-r}$, to achieve exact unlearning.
Simultaneously, this unlearning process needs to be efficient in order to respond to the user's request promptly. Finally, we aim to minimize any performance degradation after implementing the unlearning designs to ensure that users remain satisfied with the recommendation quality.

\section{Methodology}

In this section, we commence with presenting an overview of our approach, encompassing the model framework and the unlearning process. Following that, we provide a detailed discussion of the pivotal components of our method.

\subsection{Overview}
To enable exact and efficient unlearning based on retraining, our APA framework employs a partitioning strategy to train and construct the LLMRec model. Our APA framework, as illustrated in Figure~\ref{fig:framework}, encompasses three key phases:
\begin{itemize}[leftmargin=*]
    
    \item [1)] Data and Adapter Partition: We partition the training data $\mathcal{D}$ into $K$ balanced and disjoint shards, denoted as $\{\mathcal{D}{1},\dots,\mathcal{D}_{K}\}$. Given that LLM relies on text semantics for predictions, we perform the partition based on the text semantics of the samples, utilizing a K-means clustering method. Once the data shards are obtained, we proceed to train an individual LoRA adapter (a sub-adapter) for each shard in TALLRec. For the $k$-th data shard $\mathcal{D}_{k}$, we train a LoRA adapter parameterized with $\Phi_{k}$ according to Equation~\eqref{eq:tallrec} (replacing $\mathcal{D}$ and $\Phi$ in the equation with $\mathcal{D}_{k}$ and $\Phi_{k}$).
    
    \item[2)] Adapter Aggregation: At the serving stage, we perform adapter aggregation, which involves merging the weights of different LoRA adapters to create a unified adapter. We just use the aggregated  LoRA adapter for inference. Importantly, we employ a sample-adaptive aggregation strategy, tailoring the aggregated adapter to the specific sample for improved performance. This part is the key to ensuring performance and inference efficiency.  
\end{itemize}

\vspace{+5pt}
\noindent \textit{Unlearning.} When a user requests to erase data $\mathcal{D}_{r}$, only the sub-LoRA adapters affected by $\mathcal{D}_{r}$ need to be retrained, obviating the need to retrain the entire model and facilitating acceleration. In theory, we only need to invest a $\frac{|\mathcal{D}_{r}|}{K}$ cost for full retraining to achieve precise unlearning. With the support of two considerations, $\frac{|\mathcal{D}_{r}|}{K}$ can be kept at a low value, resulting in significant acceleration: Firstly, it is often assumed that user requests arrive in a streaming manner, usually, only one sample needs to be unlearned at a time. Secondly, given the few-shot learning capabilities of TALLRec (a few hundred samples are adequate to train an effective LoRA), the data partition can be fine-grained, allowing for a relatively high value of $K$.

After the LLMRec has been constructed, the unlearning process is simple and straightforward. Therefore, the essence of the process lies in our partition and aggregation stages. We now delve into the details of these two phases.

\subsection{Partition}
The partitioning phase is the key to training a LLMRec model, involving two key parts: 1) data partition, and 2) training a sub-adapter module for each partitioned data shard. We next elaborate on the two parts.

\subsubsection{Data Partition} 
For data partitioning, the crucial factor is ensuring that data within the same shard share related knowledge, creating homogeneity of knowledge within a shard and heterogeneity across shards. This aids in the effective learning of sub-adapters with limited data and their subsequent aggregation. Prior methods for recommendation unlearning relied on collaborative embedding to perform partitioning, utilizing the K-means algorithm to group samples with similar collaborative information within the same shard. Similarly, given that LLMRec relies on text semantics for prediction, we propose partitioning data based on semantics.

Our data partition method is detailed in Algorithm~\ref{alg:partiton}, consisting of three key parts. Initially, we utilize the original LLM (without fine-tuning) to derive the hidden representations (denoted as $h_{i}$) of the input instructions for each training sample $(x_i, y_i)$ in $\mathcal{D}$, capturing the text semantics (line 2). Subsequently, we employ K-means on the obtained hidden representations, resulting in $K$ clusters and $K$ clustering centers (denoted as ${a_{1}, \dots, a_{K}}$) (line 3). The clusters generated directly by $K$ can be highly unbalanced, potentially making unlearning inefficient for large shards. Therefore, we take further steps to balance the clusters (lines 4-10). Instead of directly assigning a sample to the nearest cluster, we take into account the cluster size: if the size of the closest cluster exceeds a certain threshold, we assign the sample to the nearest cluster whose size is still below that threshold. Formally, for all samples, we calculate their cosine distance to each cluster center as follows:
\begin{equation}\small
\label{cos}
    dist(h_{i},a_k) = -cos(h_{i},a_{k}).
\end{equation}
We store all distances in a list, denoted as $F=\{(x_i,y_i, dist(h_{i},a_{k})) | i\leq|\mathcal{D}|, k\leq |K| \}$, and then sort the list based on the $dist()$ function. We refer to the sorted list as $F_{s}$. Subsequently, we orderly examine each element $(x_{i'},y_{i'}, \text{dist}(h_{i'},a_{k'}))$ in $F_{s}$ to achieve balanced clustering. If $(x_i',y_i')$ has not yet been assigned to any cluster, we assign it to the $k'$-th cluster, denoted as $\mathcal{D}_{k'}$.

\subsubsection{Sub-adapter Training} After obtaining the partitioned data, we proceed to train an individual LoRA adapter for each data shard, following the approach of TALLRec. Formally, for the $k$-th data shard, the optimization objective is as follows:
\begin{equation}\small \label{eq:tallrec}
\mathop{\max}\limits_{\Phi_{k}} \sum_{(x_{i},y_{i})\in\mathcal{D}_{k}}\sum_{t=1}^{|y|}log(P_{\Theta_{0} + \Phi_{k}}(y_{i}|x_{i})),
\end{equation}
where $\Phi_{k}$ denotes the model parameters of the $k$-th LoRA for the $k$-th data shard,  $P_{\Theta_{0} + \Phi_{k}}(y_{i}|x_{i})$ denote the prediction probability of LLMRec with the $k$-th LoRA for $y_{i}$.





\subsection{Aggregation} \label{sec:aggregation}
During the serving prediction stage, aggregating knowledge from the sub-models is essential to enhance the overall prediction quality. Typically, prediction result aggregation is a commonly used approach. However, this method necessitates performing LLM inference $K$ times, as it requires computing $P_{\Theta_{0} + \Phi_{k}}(y_{i}|x_{i})$ for k=1, 2, $\dots$, $K$. To address this challenge, we introduce aggregation at the LoRA adapter model weight level, \ie adapter aggregation. This technique combines the weights of multiple LoRA adapters, creating a single LoRA adapter that allows for one-pass prediction. 


Given that a weight matrix in the original LLM corresponds to a pair of rank-decomposition weight matrices, as shown in Equation~\ref{eq:lora}, we consider two levels of aggregation:
\begin{itemize} [leftmargin=*]
    \item Decomposition level: At this level, 
    each model weight of LoRA serves as the unit for model aggregation. We directly aggregate the $A$ and $B$ matrices defined in Equation~\eqref{eq:lora} from different LoRA adapters using weight averaging. Formally, a aggregated LoRA layer can be defined as follows:
    \begin{equation}
      \begin{split} \label{eq:d-soup}
          &\bar{o} = W_{0}x + \bar{B}\bar{A}x,\\
          & \bar{B} = \sum_{k=0}^{K} \omega_{k} B_{k}, \, \bar{A} = \sum_{k=0}^{K} \omega_{k} A_{k},
      \end{split}
    \end{equation}
    where $\bar{B}$ represents the aggregated $B$ matrix,  $\bar{B}_{k}$  represents the $B$ matrix of the $k$-th sub-adapter, similarly for those of $A$; $\bar{o}$ denotes the layer output in the aggregated LoRA adapter, and $\omega_{k}$ is the aggregation weight for the $k$-th sub-adapter, where a higher value indicates higher attention. The method for assigning $\omega_{k}$ is described later. 

    \item Non-decomposition level: At this level, the weight unit of the original LLM serves as the aggregation unit for the adapter aggregation. Then, we aggregate the ``BA'' result defined in Equation~\eqref{eq:lora} from all sub-LoRA adapters using weight averaging. Formally, a aggregated LoRA layer can be formulated as follows:
    
    \begin{equation} \label{eq:nd-soup}
      \begin{split}
          &\bar{o} = W_{0}x + \overline{BA}x,\\
          & \overline{BA} = \sum_{k=0}^{K} \omega_{k} B_{k}A_{k},
      \end{split}
    \end{equation}
    where , $\overline{BA}$ represents the aggregated LoRA weights, and other symbols have the same meanings as in Equation~\eqref{eq:d-soup}.
\end{itemize}

\noindent \textbf{Sample-adaptive Strategy.} 
Different testing samples require varying levels of knowledge from different sub-models for accurate prediction, suggesting the need for an adaptive attention allocation when aggregating sub-adapters. To avoid introducing additional training and unlearning, we explore a heuristic approach to assign aggregation weights to different sub-adapters. Based on our partition, one straightforward solution is to use text semantic similarity to determine these weights, giving higher priority to the adapter corresponding to the data shards with greater similarity to the sample. However, selecting an adapter solely based on input similarity doesn't guarantee better prediction accuracy. To address this concern, we devise a method that leverages validation prediction errors to enhance the assignment mechanism. In essence, for each testing sample, we rely on the prediction errors of the most similar validation samples to assess the suitability of a particular adapter for that specific testing sample and allocate the attention weights accordingly. This approach ensures more accurate weight assignments for effective prediction.


Specifically, for each testing sample $(x, y)$, we initially identify the top-n most similar samples from the validation set, calculating the similarities similar to Equation~\eqref{cos}. These identified similar samples are denoted as $N_{v}$. Next, we measure the average prediction error among $N_{v}$ for each sub-adapter as follows:
\begin{equation}\small \label{eq:error}
    error_{k} = \frac{1}{|\mathcal{N}_{v}|} \sum_{(x_i,y_i) \in \mathcal{N}_v} error(y,P_{\Theta+\Phi_{k}}(y_i|x_i)),
\end{equation}
where $|\mathcal{N}_{v}|$ represents the size of $\mathcal{N}_{v}$, and $error_{k}$ stands for the average prediction error of the $k$-th sub-LoRA adapter. Subsequently, for this testing sample, we assign higher attention weights to the sub-LoRA adapter with lower prediction errors. Formally, the attention weight $\omega_{k}$ for the $k$-th sub-LoRA adapter is calculated as follows:
\begin{equation}\small  \label{eq:omega}
\omega_{k} = \frac{exp(\tau \cdot -error_{k}) }{\sum_{k'=0}^{K} exp(\tau \cdot -error_{k'}) },  
\end{equation}
where $\tau$ represents the temperature parameter, controlling the strength of the assignment mechanism. When $\tau=0$, the mechanism becomes ineffective, allocating equal attention weights for all sub-LoRA adapters.

\begin{algorithm}[t]
	\caption{Balanced Semantic-aware Data Partition}
	\LinesNumbered
	\label{alg:partiton}
	\KwIn{ 
 training instruction data $\mathcal{D}$, cluster number $K$, and  maximum size of each shard $t$}
 	\KwOut{The Shards $\{\mathcal{D}_{0},...,\mathcal{D}_{K}\}$}
   Initialize $\mathcal{D}_{0}$, $\dots$, $\mathcal{D}_{K}$\; 
   
   Compute hidden representation $h_{i}$ of $x_{i}$ in the original LLM for each training sample $(x_i,y_i)\in \mathcal{D}$\;

    Runing the K-means with all hidden representations $\{h_{i}|i\leq |\mathcal{D}|\}$, obtaining cluster centers: $\{a_{0},a_{1},...,a_{K}\} = K\text{-}means(\{h_{i} |i\leq |\mathcal{D}|\},K)$\;

    For each sample $(x_i,y_i)$ and each cluster center $a_{k}$, compute their cosine distance using $h_i$, \ie $dist(h_{i}, a_{k})$, storing $(x_i,y_i,dist(h_{i}, a_{k}))$ in a list $F$\;

    Sort $F$ in ascending order to get $F_{s}$\;

    \For{each $(x_i',y_i',dist(h_{i'}, a_{k'}))$ in $F_{s}$}{
    \If{$|\mathcal{D}_{k'}|<t$ and $(x_{i},y_{i})$ has not been assigned}{
    $\mathcal{D}_{k'} \leftarrow \mathcal{D}_{k'} \cup (x_{i},y_{i})$
    }
    }
    return $D$;
\end{algorithm}

\vspace{+5pt} 
\textit{Discussion.} It is worth mentioning that our method is developed specifically for LoRA-based LLMRec. However, since the PEFT method commonly incorporates adaptation modules, we can directly extend our method to LLMRec models developed using other PEFT techniques like Adapter Tuning~\cite{houlsby2019parameter}. This flexibility allows us to apply our method to a wider range of LLMRec architectures.

\section{EXPERIMENTS}
In this section, we conduct a series of experiments to answer the following research questions:
\begin{itemize}[leftmargin=*]
    \item \textbf{RQ1}: How does APA perform in terms of recommendation performance and unlearning efficiency compared to the state-of-the-art exact unlearning methods for LLMRec?
    \item \textbf{RQ2}: 
    How does APA perform in terms of inference efficiency?
    \item \textbf{RQ3}: How do different components of the proposed APA influence its effectiveness?
\end{itemize}

\begin{table*}[]
\caption{Comparison of different unlearning methods on recommendation performance, where `\textbf{APA(D)}'/`\textbf{APA(ND)}' represents APA implemented with decomposition/non-decomposition level aggregation, and $\bigtriangleup$ represents the gap between retraining and the unlearning method in terms of AUC. `Bef. Agg.' represents the average $AUC$ of the sub-model.}
\label{main}
\begin{tabular}{@{}ccccccc@{}}
\toprule
\textbf{Book}  & \textbf{Retraining} & \textbf{SISA} & \textbf{GraphEraser} & \textbf{RecEraser} & \textbf{APA(D)} & \textbf{APA(ND)} \\ \midrule
\textbf{Bef. Agg.}   & -           & 0.6570        & 0.6443                & 0.6620             & 0.6578
                 & 0.6578                \\\hdashline
\textbf{AUC}   & 0.6738           & 0.6728        & 0.6684                & 0.6732             & 0.6829                 & 0.6846                \\ 
\textbf{$\bigtriangleup$}  & -               & -0.001        & -0.0052                & -0.0006             & 0.0091                      & 0.0108                 \\
\midrule
\textbf{Movie} & \textbf{Retraining} & \textbf{SISA} & \textbf{GraphEraser} & \textbf{RecEraser} & \textbf{APA(D)} & \textbf{APA(ND)} \\ \midrule
\textbf{Bef. Agg.}   & -           & 0.7003        & 0.6672               & 0.6712            & 0.6696               & 0.6696                \\\hdashline
\textbf{AUC}   & 0.7428           & 0.7035        & 0.6903                & 0.6937             & 0.7259                 & 0.7256                \\
\textbf{$\bigtriangleup$}  & -                & -0.0393        & -0.0525                & -0.0491              & -0.0169                 & -0.0172                \\ \bottomrule
\end{tabular}
\end{table*}

\subsection{Experimental Settings}
\subsubsection{Datasets}
We conduct experiments on two distinct real-world datasets, which are widely recognized and used within the realm of recommender systems:
\begin{itemize}[leftmargin=*]
\item \textbf{Book}. We utilize the BookCrossing dataset~\cite{bookcross}, which consists of user ratings ranging from 1 to 10. This dataset also provides textual descriptions of books, including attributes like ``book author'' and ``book title''. Furthermore, we binarize the ratings based on a threshold of 5. That is, ratings exceeding 5 are considered as ``like'', while ratings below 5 are labeled as ``dislike''.
    \item \textbf{Movie}. We utilize the MovieLens100K~\cite{movielens} benchmark dataset, which comprises user ratings ranging from 1 to 5, following ~\cite{lightgcn,zhang2023reformulating}, we treat the ratings as ``like’’ (corresponding to the "Yes" in the TALLRec instruction out) if the ratings are higher than 3 and otherwise ``dislike’’. The dataset includes comprehensive textual attributes of the movies, such as ``title’’ and ``director''. 

\end{itemize}
We strictly adhere to the pre-processing procedures outlined in the TALLRec paper~\cite{tallrec} for data filtering, data splitting, and instruction data construction for both datasets. In particular, considering TALLRec's capability to efficiently learn recommendations and yield good performance with a minimal number of training samples, we constrain our training size to 1024 (larger than the maximum size of 256 mentioned in the TALLRec paper). Similar to the TALLRec setting, the validation set comprises 500 samples for both Movie and Book, while the testing consists of 1000 samples for both the Movie and Book datasets.


\subsubsection{Compared Methods}
We focus on exact unlearning, but there is currently no specific work designed for LLM. Therefore, to serve as a baseline, we consider extending the following traditional exact unlearning baselines to TALLRec:
\begin{itemize}[leftmargin=*]
    \item \textbf{Retraining} 
    This represents the straightforward retraining approach, \ie retraining the entire model from scratch while excluding the unusable data. We implement it by retraining TALLRec from scratch, excluding the unusable data. This method serves as the gold standard in terms of recommendation performance. 
    
    
    \item \textbf{SISA~\cite{sisa}} is the earliest known partition-enhanced retraining method. It randomly divides data and aggregates sub-model predictions through methods such as averaging or majority voting. We extend this approach to TALLRec, employing its average-based aggregation.
    
    
    \item \textbf{RecEraser~\cite{receraser}} is a recommendation-specific unlearning method, sharing similarities with SISA but incorporating unique partitioning strategies to preserve collaborative information. We adapt it for LLMRec based on its UBP version. Notably, its prediction aggregation involves training with $K$ TALLRec, requiring overmuch computational resources. In our adaptation, we directly replace it with the aggregation strategy of SISA.
    
    
    \item \textbf{GraphEraser~\cite{graphunlearn}} is an unlearning method designed for graph-structured data (including the bipartite graph structure of interaction data). It employs node clustering techniques, namely BEKM and BLPA, for graph data partitioning. We extend GraphEraser to TALLRec using the BEKM-based partition. Similar to RecEraser, we adopt the aggregation strategy of SISA for it.
    
\end{itemize}
Regarding our APA, we implement two versions using different levels of aggregation, as defined in Section~\ref{sec:aggregation}. We denote the version with decomposition-level aggregation as APA(D) and the version with non-decomposition-level aggregation as APA(ND).


\subsubsection{Evaluation Setting}
Our objective is to achieve precise and efficient unlearning for LLMRec while preserving recommendation performance. Therefore, our evaluation focuses on three aspects: 1) the completeness of data removal, 2) unlearning efficiency, and 3) recommendation performance. Since all compared methods are built on retraining (from scratch) without the unusable data, the first aspect is inherently maintained. Following the RecEraser paper~\cite{receraser}, we do not consider this aspect for evaluation. To assess recommendation performance, we use the Area under the ROC Curve (AUC) metric, following the TALLRec paper. Additionally, we introduce another performance loss metric $\Delta$ to measure the recommendation performance loss of a method relative to the Retraining method. This metric is calculated by the difference in AUC between the method and the Retraining method, with higher values indicating less performance loss. For evaluating unlearning efficiency, we directly utilize the unlearning time (retraining time). Additionally, considering the inference cost for LLMs, we further compare the inference time.



\subsubsection{Implementation Details}

As all methods utilize TALLRec as the backbone recommendation model, we apply the same hyper-parameter settings for them to learn the recommendation model, following the original configuration outlined in the TALLRec paper. Concerning the unlearning setting, for all partition-based base models, we set the shard size to 256 for the data partition, resulting in $K=4$ training data shards. For the specific hyper-parameters of the baselines, we tune them in accordance with the settings provided in the original papers, whenever available for our extension. Regarding the proposed APA, we set $\tau$ (in Equation~\eqref{eq:omega}) to 1000 for both dataset. For the neighbor size $|\mathcal{N}_{v}|$ in Equation~\eqref{eq:error}, we set it to 20 for Movie and 100 for Book. All these hyperparameters are tuned on the validation set, and all experiments are conducted on the same machine equipped with NVIDIA A40 GPUs.

\subsection{Main Results (\textbf{RQ1})}
In this subsection, we evaluate all unlearning methods based on two criteria: recommendation performance and unlearning efficiency.  It is essential to note that all compared methods inherently achieve complete removal of unusable data; therefore, we omit the comparison in the aspect of exact unlearning~\cite{receraser}.

\subsubsection{Accuracy Comparison}
We compare the accuracy of APA with that of the baselines to assess its ability to maintain recommendation performance during unlearning. Higher ability is indicated by less performance loss compared to the Retraining method. The comparison results are summarized in Table~\ref{main}, where we additionally include the averaged performance of the sub-models for the partition-based methods and draw the following observations:
\begin{itemize}[leftmargin=*]
    \item  
    All methods with aggregation demonstrate improved $AUC$ compared to the averaged $AUC$ of their corresponding sub-models. This underscores the significance of aggregating knowledge from sub-models to enhance performance.

    \item Compared to the baselines, APA exhibits less performance loss compared to the reference Retraining method and can even bring improvements. These results highlight the superior ability of our method to maintain recommendation performance during unlearning. This superiority can be attributed to the compatibility between our partitioning method and aggregation, as well as the adaptive aggregation approach based on validation performance, which pays more attention to high-performance sub-adapters.
    
    \item 
     In contrast, SISA, RecEraser, and GraphEraser show much inferior performance compared to the reference method Retraining. Particularly on the movie dataset, these baselines exhibit a significant decline in recommendation performance. This suggests that the direct application of traditional methods to TALLRec results in a substantial compromise in recommendation performance.

    \item 
    The two versions of APA with different levels of adapter aggregation (APA(D) and APA(ND)) demonstrate similar performance. This indicates that treating the LoRA rank-decomposed parameter as the aggregation unit or the original LLM parameter unit as the aggregation unit does not affect the effectiveness of our adaptive aggregation method.

\end{itemize}

\subsubsection{Unlearning Efficiency Comparison}
\begin{figure}
    \centering
    \subfigure[\textbf{Unlearning time}]{\label{fig:unlearntime}\includegraphics[width=0.22\textwidth]{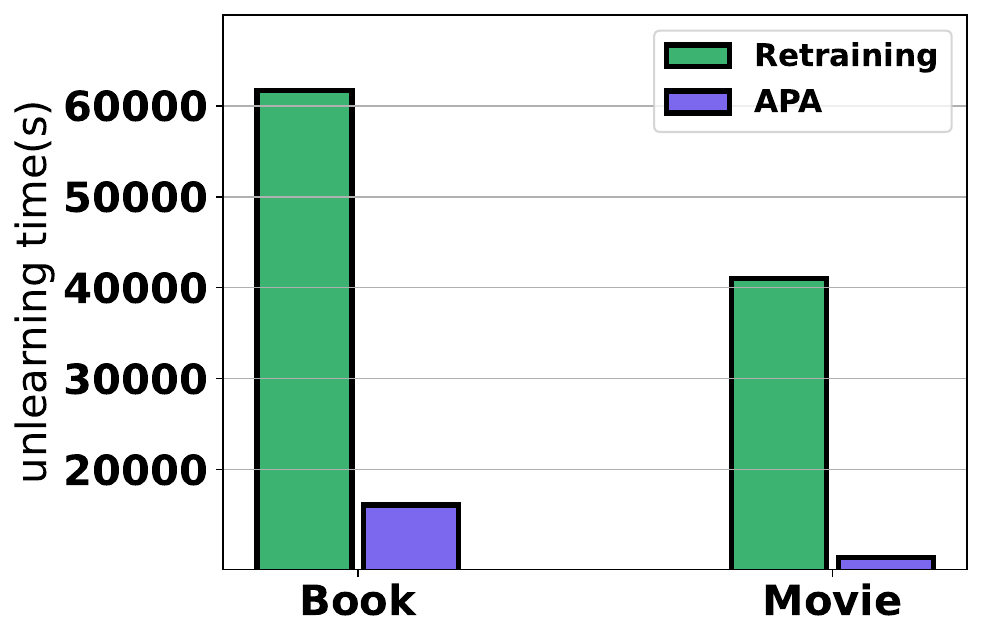}}
    \subfigure[ \textbf{Inference time}]{\label{fig:infre} \includegraphics[width=0.22\textwidth]{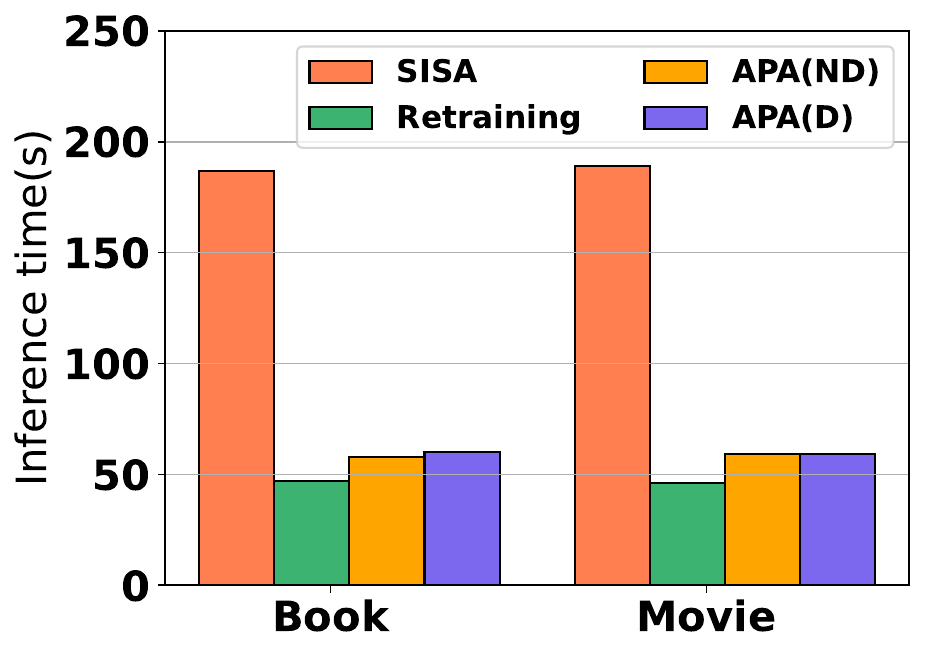}}   
    \caption{(a) Unlearning time of Retraining and APA. (b) Inference time of Retraining, SISA, APA(D), and APA(ND).}
    \vspace{-15pt}
    \label{fig:time}
\end{figure}

We next conduct experiments to explore the unlearning efficiency of our APA.
We fully follow the efficiency evaluation experiment setting in the RecEraser paper~\cite{receraser}, ensuring that only one sub-model needs to be retrained for unlearning each time. We primarily compare our method with the Retraining method, as other baselines theoretically have similar unlearning efficiency costs to us. The results are shown in Figure~\ref{fig:unlearntime}. The results demonstrate that APA significantly improves unlearning efficiency. For example, on the movie dataset, APA only took 10,335 seconds, making it 3.96 times faster (approximately $={K}$) than the Retraining method. The Retraining method is time-consuming as it is trained on the whole dataset. In contrast, APA only requires retraining the specific sub-model responsible for the unlearned data. 
Moreover, when the training data is large, we can keep a small shard size to allow for a large number of data shards $K$, considering that TALLRec can effectively learn recommendations with few samples. In this case, APA could achieve greater acceleration as long as only a few sub-adapters are affected by unusable data.

\subsection{Inference Time Comparison (\textbf{RQ2})}

In the previous section, we explored how our APA method can significantly reduce time during the unlearning process. In this section, we investigate whether APA can improve efficiency during the inference stage compared to baselines. We randomly selected 500 samples from the testing set and measured the total inference time for these samples, ensuring that only one LLM inference could be executed at a time. We compare our APA with SISA and Retraining (we omit other baselines due to their similar costs to SISA).
The experimental results are presented in Figure~\ref{fig:infre}. From the results, we have the following observations: 1) Our APA method exhibits small gaps in time efficiency compared to a single model inference (Retraining), and the delay for each sample is just approximately 0.02 seconds, which is entirely acceptable in real-world scenarios. 2) SISA has much higher inference time costs compared to Retraining and APA. This is because SISA performs prediction-level aggregation for sub-adapters, which involves additional time for inference cost. These results demonstrate the effectiveness of APA's aggregation designs in enhancing inference efficiency.

\begin{figure}
    \centering
    \subfigure[\textbf{Book}]{\label{fig:a}\includegraphics[width=0.22\textwidth]{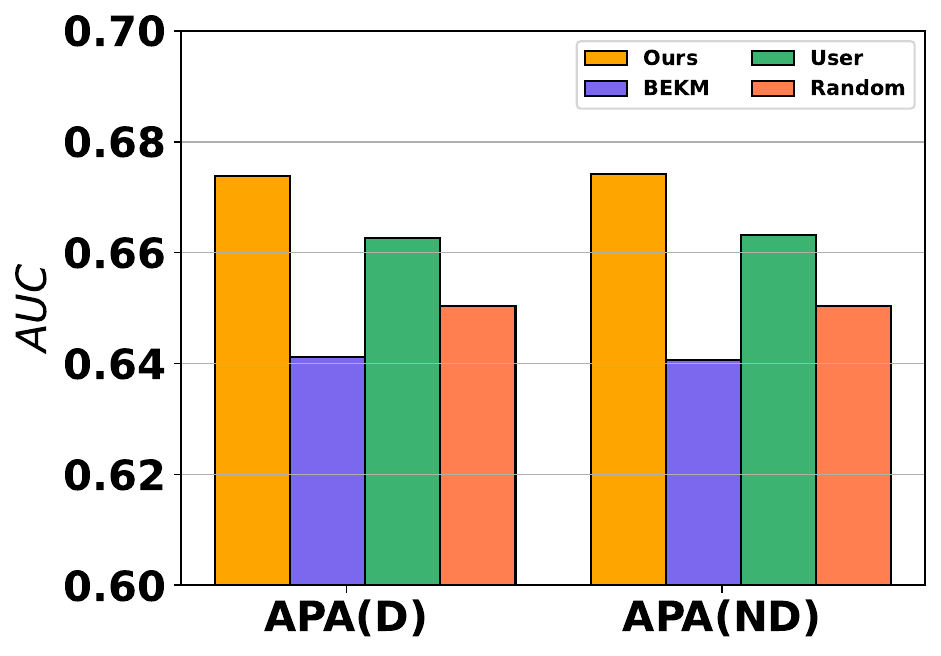}}
    \subfigure[ \textbf{Movie}]{\label{fig:b} \includegraphics[width=0.22\textwidth]{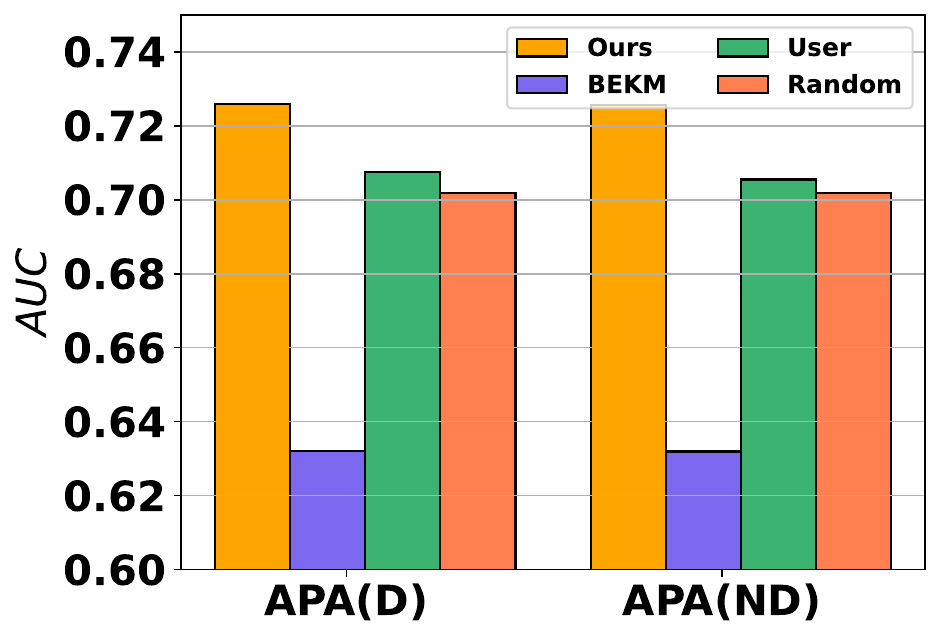}}
    \caption{Performance comparison of different data partition methods on Book and Movie datasets.}
    \label{fig:data-partition}
\end{figure}

\subsection{In-depth Studies(\textbf{RQ3})}
\subsubsection{Effect of the Data Partition Methods}

To validate the effectiveness of our proposed data partition method, we compare it with three other grouping methods: random partition, the user-based partition of RecEraser~\cite{receraser}, and BEKM partition~\cite{graphunlearn}, denoted as 'Random,' 'User,' and 'BEKM', respectively. We replace the original partition method with the three methods in our APA, respectively, and then compare their performances with the original one. The experimental findings are illustrated in Figure~\ref{fig:infre}. Based on the results, we draw the following observations: 1) Replacing the original semantic-aware method with any of the three methods would result in a performance decrease. For example, On the Book dataset, the original semantic-aware method achieves an AUC score of 0.6829, while the corresponding results were just 0.6503, 0.6627, and 0.6411 for random partition, user-based partition, and BEKM, respectively. The result shows the importance of leveraging semantics to partition for LLMs, which could ensure better heterogeneity of data shards to facilitate better aggregation for enhancing recommendation performance.


\begin{figure}
    \centering
    \subfigure[\textbf{Book}]{\label{fig:a}\includegraphics[width=0.22\textwidth]{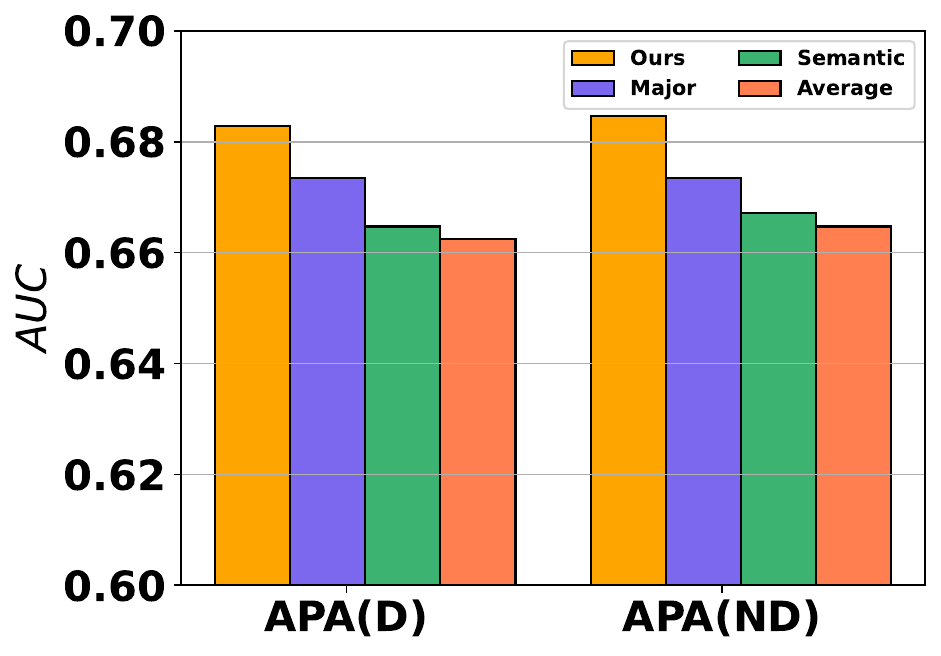}}
    \subfigure[ \textbf{Movie}]{\label{fig:b} \includegraphics[width=0.22\textwidth]{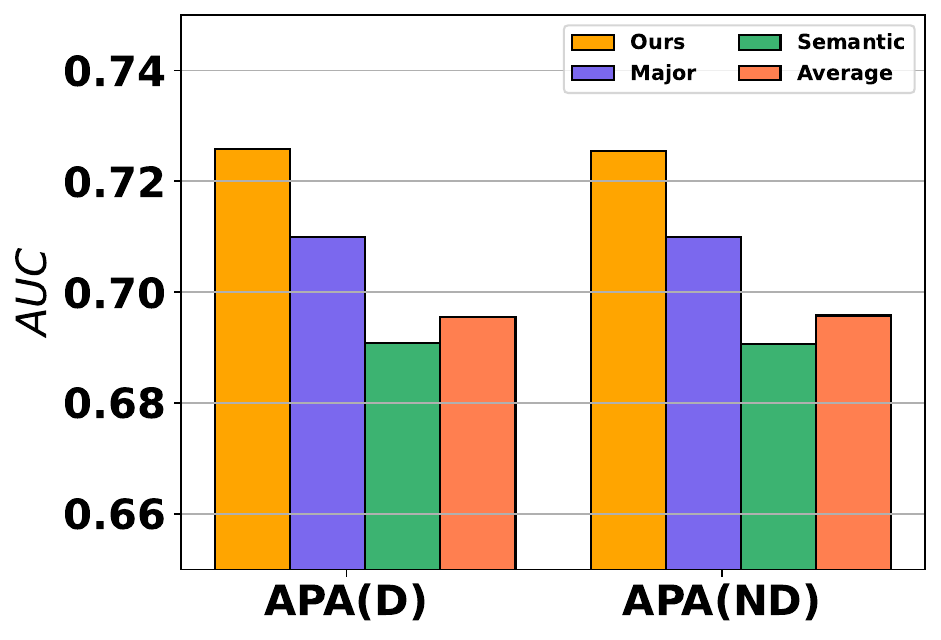}}
    \caption{Performance comparison of different aggregation weight assignment methods on Book and Movie datasets.}
    \vspace{-15pt}
    \label{fig:agg}
\end{figure}

\subsubsection{Effect of the Sample-Adaptive Method}

We proceed to assess the model utility of different aggregation weight assignment methods to demonstrate the effectiveness of our proposed sample-adaptive method. We compare our methods with the following three choices: 1) average-based, assigning equal weight for each sub-adapter, 2) major-based, assigning all weights to the one with the highest $\omega_{k}$ computed by our method, 3) semantic-based, which assigns weight according to the semantic similarity of the sample to the center of different shards. We compare the APA implemented with our assignment method with the variants of APA implemented with the three methods. The experimental results are presented in Figure~\ref{fig:agg}, with 'Average,' 'Major,' and 'Semantic' denoting the compared three choices, respectively.
Here are some observations we found: 1) The average-based method underperforms our sample-adaptive method on both datasets, highlighting the importance of assigning different weights for different adapters; 2) The semantic-based method also exhibits worse recommendation performance than our original method, confirming the effectiveness of utilizing validation performance information; 3) Using only the best sub-model choice by our weight assignments can maintain relatively high recommendation performance, but there is still a gap compared to our method, as shown by the results of the major-based method. These results emphasize the effectiveness of our weight assignments, and meanwhile, the importance of aggregating knowledge from different sub-adapters.

\subsubsection{Impact of the Shard Size}

\begin{figure}[t]
  \centering
  \subfigure[\textbf{Recommendation Performance}]{\label{fig:shardinf}\includegraphics[width=0.23\textwidth]{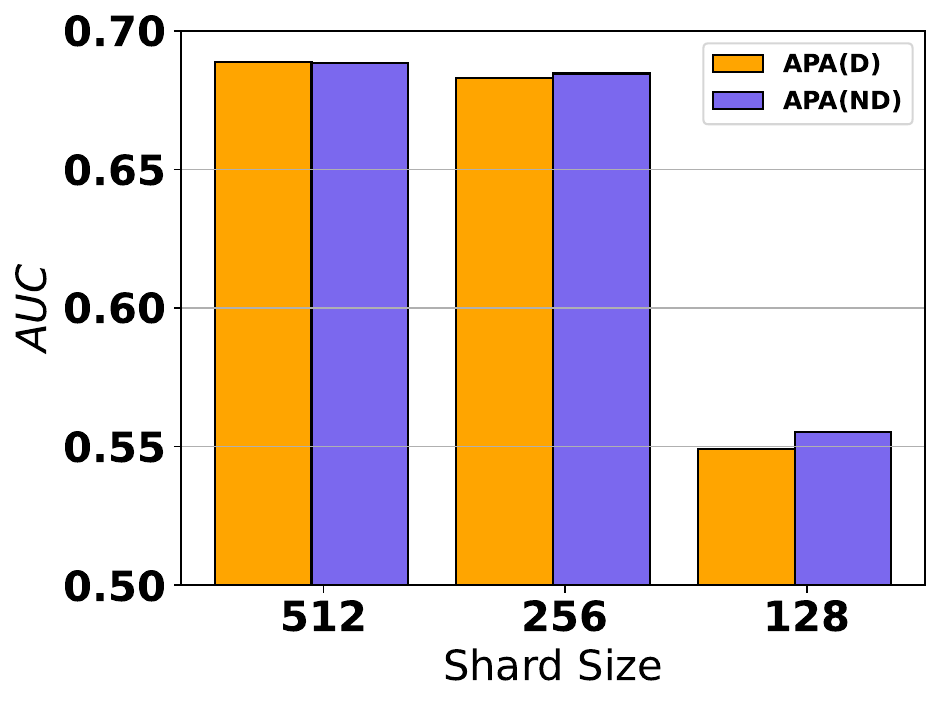}}
    \subfigure[ \textbf{Unlearning Time}]{\label{fig:shardunlearn} \includegraphics[width=0.23\textwidth]{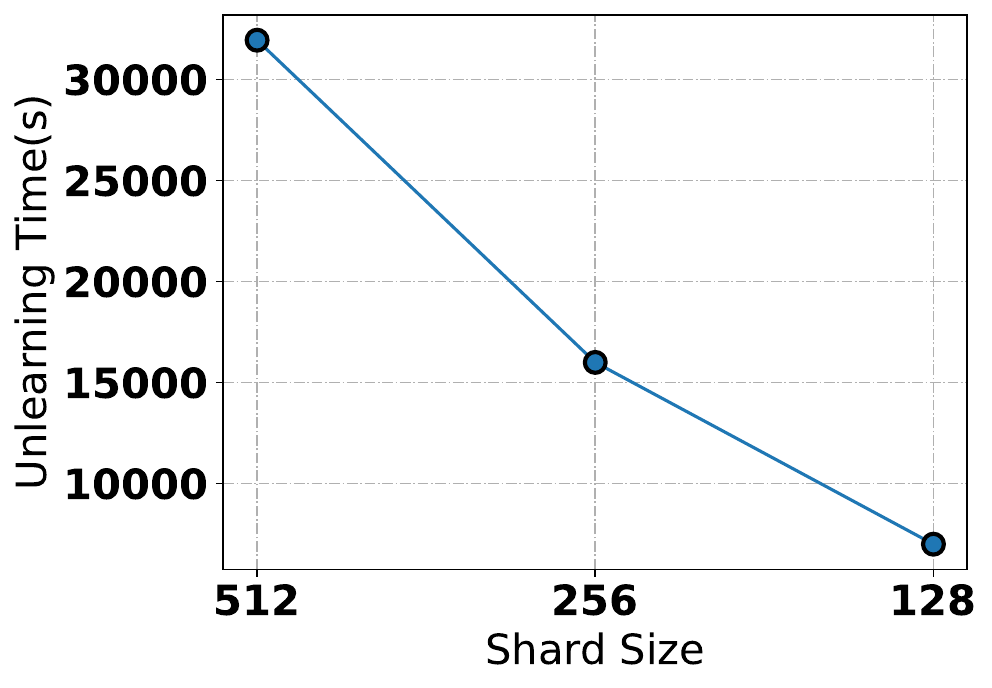}}
  \caption{Impact of the shard size on the unlearning efficiency and performance on Book dataset. (a) shows the recommendation performance and (b) shows the unlearning time cost.}
  \label{fig:shard}
\end{figure}

We investigate the influence of the shard size on the Book dataset. We configure the shrad size as $\{512, 256, 128\}$ and evaluate unlearning efficiency and recommendation performance. The experimental results are displayed in Figure~\ref{fig:shard}. We find that 1) as the shard size decreases, the unlearning time significantly reduces; 2) In terms of recommendation performance, as the shard size decreases, it remains relatively stable before decreasing. For example, when the shard size is 512 and 256, the performance of APA remains very close, and it only significantly decreases at 128. This indicates that within a certain range, we can improve unlearning efficiency by reducing the data shard size (increasing the number of shards) while maintaining comparable recommendation performance. In this way, on the one hand, the cost of retraining individual shards decreases, and on the other hand, increasing the number of shards may keep the proportion of adapters requiring retraining relatively low, thereby enhancing unlearning efficiency.

\section{Related work}
\subsection{Machine Unlearning}
\noindent$\bullet$ \textbf{Machine Unlearning}.
Machine unlearning, the process of removing partial training data information from trained machine learning models, is essential in various domains, including recommendation, for reasons such as privacy and security concerns~\cite{cao2015towards,marchant2022hard}. This concept is known as machine unlearning~\cite{sisa}. In traditional machine learning, two main technique lines for unlearning have emerged: approximate unlearning and exact unlearning~\cite{unlearninSurvey,anothersurvey}. Approximate unlearning aims for unlearning without retraining, using techniques like influence functions~\cite{ifru,izzo2021approximate} and data augmentation~\cite{shan2020fawkes,tarun2023fast} for extreme efficiency, but it often involves incomplete removal of the data. On the other hand, exact unlearning~\cite{sisa} typically involves retraining, ensuring complete unlearning but in a time-costly manner. Existing work, like SISA~\cite{yan2022arcane, qian2022patient, receraser, graphunlearn}, focuses on partition strategies, building individual sub-models for partitioned training data shards to retrain only partial sub-models. Our method, while also based on the partition strategy, addresses new challenges posed by the large scale and high inference cost of Large Language Models (LLM). This makes our work distinct from existing methods.

\noindent$\bullet$ \textbf{LLM Unlearning}. 
The challenges presented by Large Language Models (LLMs), particularly their large scale, bring forth new considerations for unlearning. Previous efforts~\cite{HarryUnlearning,InContextUnlearn,yao2023large} have explored unlearning for LLMs, but they often involve approximate methods. For instance,  \citep{HarryUnlearning} simulates data labels to approximate the next-token predictions of a model that has not been trained on the unusable data, and then fine-tune LLM on these simulated labels for unlearning. \cite{InContextUnlearn} proposes "In Context Unlearning", which leverages in-context learning by flipping labels of unusable data to achieve approximate unlearning.  \citep{yao2023large} leverage the gradient ascent to erase the influence of unusable data on a trained model with fine-tuning. However, these methods do not achieve complete removal of unusable data and are not tailored for LLMs in the context of recommender systems. In contrast, our approach focuses on LLMRec and strives for exact unlearning, considering the significant impact of incomplete removal of sensitive data.

\subsection{Model Aggregation in LLM}
To aggregate the different models, there are two strategies: 1) output aggregation and 2) model weight aggregation. Output aggregation has been widely studied, and applied for aggregation process for partition-based unlearning, but could introduce inefficiency for LLM.  Regarding the model weight aggregation, existing work focuses on leveraging it to better finish tasks like image classification, multi-domain learning~\cite{diao2023mixture}, Cross-lingual information extraction~\cite{wang2023mt4crossoie}, etc. To our knowledge, we are the first to leverage it for the unlearning task. Meanwhile, from the technical view, our method has significant differences from existing work on the aggregation weight assignment. To achieve weight assignment, previous works usually considered 1) average aggregation~\cite{modelsoup}, 2) greedy aggregation, and 3) learning-based aggregation like simulating Mixture-of-Experts (MoE)~\cite{wang2023mt4crossoie,diao2023mixture}. Differently, we innovatively assign weights to different sub-models based on the prediction quality of similar verification samples, which can achieve effective adaptive weight assignments without learning.

\section{CONCLUSION}
In this study, we introduce a novel and efficient APA framework, which, to the best of our knowledge, is the first exact unlearning method designed for large language model-based recommendation (LLMRec). To achieve efficient unlearning while preserving high recommendation performance, we propose a data partition method based on text semantics. Additionally, we employ parameter-level adapter aggregation to create an aggregated adapter to mitigate the high inference cost associated with traditional methods. We carry out comprehensive experiments on two real-world datasets, offering insightful analysis of the effectiveness and efficiency of our approach in removing interaction data. 

In our future endeavors, we aim to expand our approach to encompass other PEFT methods, thus widening the adaptability of our method across diverse LLMRec architectures. Moreover, we are exploring methods to enhance unlearning efficiency in batch settings, enabling the management of more unlearn data.

\bibliographystyle{ACM-Reference-Format}
\bibliography{sample-base}
\balance
\appendix




\end{document}